\documentclass[3p,times]{elsarticle}

\usepackage{amssymb}

\usepackage[figuresright]{rotating}
\usepackage{graphicx}
\usepackage{caption}
\usepackage{subcaption}
\usepackage{amsmath}
\usepackage{color}
\usepackage{xcolor}
\usepackage[normalem]{ulem}

\begin{document}

\begin{frontmatter}




\title{Polish Doughnuts around Scalarized Kerr Black Holes}


\author[Oldenburg]{M. C. Teodoro}
\author[Tuebingen]{L. G. Collodel}
\author[Tuebingen,Sofia3]{D.  Doneva}
\author[Oldenburg]{J. Kunz}
\author[Sofia]{P. Nedkova}
\author[Tuebingen,Sofia,Sofia2]{S. Yazadjiev}
\address[Oldenburg]{Institute of Physics, University of Oldenburg, 26111 Oldenburg, Germany}
\address[Tuebingen]{Theoretical Astrophysics, University of T\"ubingen, 72076 T\"ubingen, Germany}
\address[Sofia3]{INRNE, Bulgarian Academy of Sciences, 1784 Sofia, Bulgaria}
\address[Sofia]
{Department of Theoretical Physics, Sofia University, Sofia 1164, Bulgaria}
\address[Sofia2]{Institute of Mathematics and Informatics, Bulgarian Academy of Sciences, Acad. G. Bonchev St. 8, Sofia 1113, Bulgaria}

\begin{abstract}
In this work we aim to investigate non-mainstream thick tori configurations around Kerr Black Holes with Scalar Hair  (KBHsSH). 
For that goal, we provide a first approach using constant specific angular momentum non-self-gravitating Polish doughnuts.
Through a series of examples, we show the feasibility of new topologies, such as double-centered tori with two cusps as well as similar structures as the ones found for rotating Boson Stars (BSs), namely tori endowed with two centers and a single cusp.
These KBHsSH' solutions are also shown to possibly house static surfaces, associated to the static rings present in these spacetimes. 
Through this report we highlight the differences between these fluid configurations when housed by some KBHsSH examples, standard Kerr black holes and rotating BSs.
\end{abstract}


\end{frontmatter}


\section{Introduction}
\label{Introduction}

Commonly used as a first approach to thick accretion disk theory around compact objects, Polish doughnuts with a constant specific angular momentum distribution are a simple but yet powerful model. 
They are simple in the sense that their construction requires nothing but the assumption of a non-self-gravitating fluid circularly rotating a compact central object. 
On the other hand this model is powerful for it provides valuable insights on the general topology of thick accretion disks and can also be used as initial condition for simulations.

The subject of accretion disks is of ever-growing interest, for these structures, through the transformation of potential gravitational energy into
radiation, can provide insights for the emission of X-ray binaries, active galactic nuclei and quasars. 
A good example of the usefulness of thick accretion disk models is the imaging of the compact object at the center of M87 \cite{EventHorizonTelescope:2019dse,EventHorizonTelescope:2019uob,EventHorizonTelescope:2019jan,EventHorizonTelescope:2019ths,EventHorizonTelescope:2019pgp,EventHorizonTelescope:2019ggy}.

In fact, in order to process the data from the Event Horizon Telescope Collaboration (EHTC), general relativistic magnetohydrodynamics (GRMHD) simulations of disks were necessary. 
In this context, Polish doughnuts are often used as an initial condition for such type of simulations. 
GRMHD simulations with a similar setup are also used to address advection-dominated flows, the evolution of weakly magnetized disks, jet formation and the differences between possible images of black holes and boson stars (BS) \cite{Narayan:2012yp,McKinney:2004ka,McKinney:2006tf,Vincent:2015xta,Olivares:2018abq,Vincent:2020dij}. 
Outside the context of GRMHD simulations, Polish doughnuts are also used to address possible tori geometries around exotic spacetimes such as Kerr-de Sitter backgrounds, distorted static BHs, deformed compact objects, Kehagias-Sfetsos naked singularities and BSs \cite{Slany:2005vd,Gimeno-Soler:2018pjd,Faraji:2020xzo,Faraji:2020tmv,Stuchlik:2014jua,Meliani:2015zta,Teodoro:2020kok}. 
Although some models of doughnuts may include magnetic fields and non-constant specific angular momentum distribution, the constant specific angular momentum case remains being an important first step in this field of research, since in this case the torus solutions are marginally stable for Kerr BHs and hold similar topologies as the ones found for different specific angular momentum distributions \cite{1975ApJ...197..745S,1980AcA....30....1J}.

The realization of KBHsSH, i.e.~equilibrium configurations of spinning black holes surrounded by solitons, was first reported a few years ago in \cite{PhysRevLett.112.221101,Herdeiro_2015}.
These are hairy black holes that continuously connect Kerr BHs to regular, rotating BSs. They emerge from a superradiant instability that causes an initial cloud of scalar field to grow and eventually reach the equilibrium found in the stationary case, where the solitonic hair coexists with the hole in a bound system with synchronized motion. 
Since then, several generalizations have been made, such as adding self-interactions to the scalar sector  \cite{PhysRevD.92.084059,PhysRevD.103.104029,Herdeiro:2016gxs}, considering higher winding numbers and excited states \cite{DELGADO2019436,PhysRevD.99.064036}, as well as admitting different field theories for the matter sector such as the Proca hair \cite{Herdeiro_2016}, and different theories for the gravitational sector such as  tensor-scalar and -multiscalar theories  \cite{Kleihaus:2015iea,PhysRevD.102.084032}. 
Different applications of astrophysical relevance have been so far entertained, that could help probe the existence of these objects and help better constrain global parameters through future observations. 
These include the studies of shadows, horizon geometry, Iron K$\alpha$ line, thin accretion disks and Polish doughnuts \cite{Cunha:2015yba,PhysRevD.97.124012,Ni_2016,Collodel_2021,Gimeno-Soler:2018pjd}.

As shown in \cite{Teodoro:2020kok}, rotating BSs endowed with a static ring \cite{Collodel:2017end} are able to shelter tori presenting the so called {\it static surfaces} as well as compelling topologies (see also \cite{Meliani:2015zta}). 
Similar results can be found for KBHsSH.
Double-centered tori and static surfaces are again observed, while new features such as tori endowed with two cusps can also be found. 
The aim of our paper is to show and analyze such features. 

In section 2 we discuss how the KBHsSH solutions are obtained, while in section 3 we provide the general recipe for the building of the constant specific angular momentum tori around these geometries.
Section 4 is dedicated to the displaying of a few torus examples around two chosen KBHsSH. Finally we report our conclusions in section 5.

\section{Scalarized Kerr Black Holes}

KBHsSH are simply the realization of the combined system of a black hole with self-gravitating solitonic hair, still within general relativity. 
The action that gives rise to it is therefore the Einstein-Hilbert action plus the contributions of the solitonic sector, given by a complex scalar field theory minimally coupled to gravity,
\begin{equation}
\label{eq:action}
S=\int \left[\frac{R}{2}-\frac{1}{2}g^{\mu\nu}\left(\partial_\mu\Phi^*\partial_\nu\Phi+\partial_\mu\Phi\partial_\nu\Phi^*\right)-U(\Phi)\right]\sqrt{-g}d^4x, 
\end{equation}
where $R$ is the Ricci scalar, $\Phi$ is the complex scalar field, $U$ is the scalar field potential that contains the mass term and possibly self-interaction terms, and $g$ is the metric determinant.

The equations are obtained by varying (\ref{eq:action}) with respect to the inverse of the metric, yielding non-vacuum Einstein field equations, and the real and imaginary components of $\Phi$, which gives two Einstein-Klein-Gordon equations. 
The scalar field theory is endowed with a global $U(1)$ symmetry since it is invariant under the transformation $\Phi\rightarrow e^{i\alpha}\Phi$ for any constant $\alpha$. 
Hence, a conserved Noether current arises
\begin{equation}
\label{eq:NoetherCurrent}
j^\mu=-i\left(\Phi^*\partial^\mu\Phi-\Phi\partial^\mu\Phi^*\right).
\end{equation}

KBHsSH are the outcome of a superradiant instability, when the system is rotating. 
We adopt the following metric parametrization for the line element,
\begin{equation}
\label{eq:ds2}
ds^2=-\mathcal{N}e^{2F_0}dt^2+e^{2F_1}\left(\frac{dr^2}{\mathcal{N}}+d\theta^2\right)+e^{2F_2}r^2\sin^2\theta\left(d\varphi-\omega dt\right)^2,
\end{equation}
where $\mathcal{N}=1-r_H/r$, with horizon radius $r_H$ in this coordinate system. 
The four metric functions we need to solve for, \{$F_0,F_1,F_2,\omega$\}, are all dependent on both $r$ and $\theta$. 
The spacetime is therefore stationary and axisymmetric and possesses two Killing vectors associated with these isometries, $\xi^\mu=\vec{\partial}_t$ and $\chi^\mu=\vec{\partial}_\varphi$.

The complex scalar field, as opposed to the metric functions, must depend on all four spacetime coordinates to grant stability and rotation. 
Since the spacetime is stationary and axisymmetric, the Lagrangian and all equations of motion must be independent of $t$ and $\varphi$, which is only possible if $\Phi$ depends on them harmonically, in explicit form, as given by
\begin{equation}
\label{eq:PhiAnsatz}
\Phi=\phi(r,\theta)e^{i(\omega_st+k\varphi)},
\end{equation}
where $\omega_s$ is the frequency of the scalar field and $k$ is the winding number which must be an integer due to the identification $\Phi(\varphi=0)=\Phi(\varphi=2\pi)$. 
Bound states of KBHsSH in equilibrium configuration occur when the axial phase velocity of the scalar field matches the angular velocity of the horizon,
\begin{equation}
\label{eq:sync}
\frac{\omega_s}{k}=-\omega_H,
\end{equation}
i.e.~the hair is \emph{synchronized}. 
Note that this implies the no flux condition $K^\mu\partial_\mu\Phi|_H=0$, where $K^\mu\equiv\xi^\mu+\omega\chi^\mu$ is the Killing vector that defines the null hypersurface.

\begin{figure}[h]
\begin{center}
    
\includegraphics[scale=0.45]{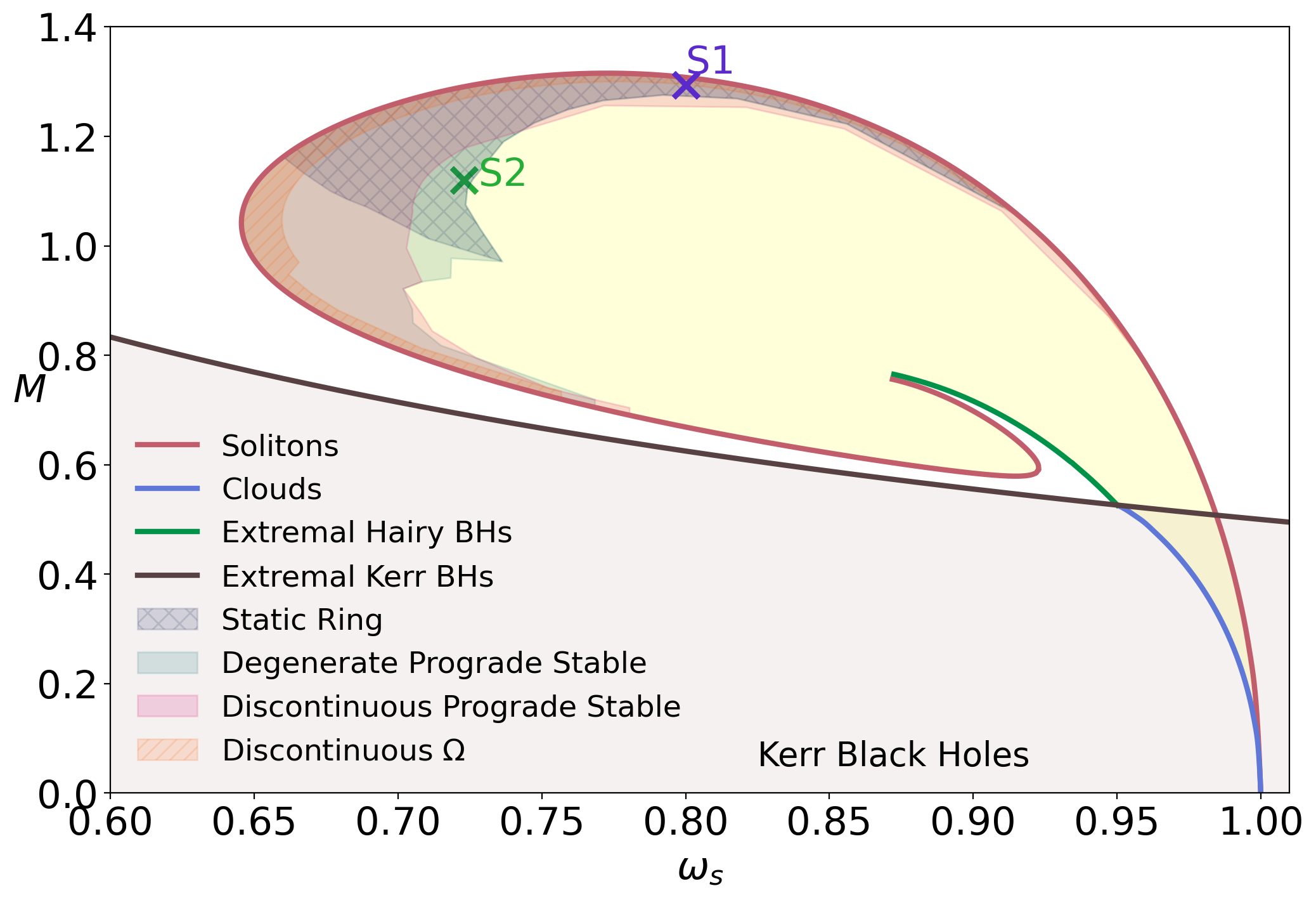}
  \caption{Domain of existence of the KBHsSH, where the selected solutions S1 and S2 are marked.}
  \label{Fig:solutions}

\end{center}
\end{figure}

In this work, we consider a non-self-interacting scalar field, so that $U=\mu^2\Phi\Phi^*/2$. 
The total mass, frequency and radial coordinate are rescaled according to the field's mass $\mu$: $M\rightarrow M\mu$, $\omega_s\rightarrow\omega_s/\mu$ and $r\rightarrow r\mu$. 
In Fig.~\ref{Fig:solutions} we display the domain of existence of KBHsSH in an $\omega_s$ vs.~$M$ parameter space, which contains three continuously connected boundaries.
The red curve represents purely solitonic solutions which are regular everywhere.
The blue curve represents Kerr BHs with a stationary solitonic cloud which is not backreacting. The green line corresponds to extremal KBHsSH. 
In this diagram, we map out some peculiar regions according to special features of the orbital parameters of a test particle in circular geodesic motion on the equatorial plane. 
The static ring, as already discussed, corresponds to solutions for which one eigenvalue of the circular orbit angular velocity goes to zero at a certain value of the coordinate $r$. 
The lowermost boundary of this area defines the occurrence of an ergoregion where the particle would otherwise have been able to stay at rest. 
The green region highlights solutions in which, for some values of $r$, both eigenvalues of the angular velocity are positive (hence just prograde orbits are defined at those points) and furthermore both orbits are stable. 
The pink domain encompasses solutions in which the stability of the prograde orbits changes more than once with varying $r$, giving rise to disjoint intervals where these orbits are stable. 
Finally, the hatched orange portion indicates spacetimes that contain regions where the angular velocity is not defined (regardless of stability) and where therefore no inertial circular orbits are allowed. 
This is an entirely different phenomenon from orbits not being defined for falling out of the light-cone, however, and non-inertial circular motion is still allowed in the respective regions. 
These orbital properties have been discussed in \cite{Collodel:2017end,Collodel_2021,Delgado:2021jxd}.

\section{Constant specific angular momentum Polish doughnuts}

The non-self gravitating axisymmetric stationary thick tori we shall present in this paper, inspired by the formalism regarding the study of rotating fluid masses in general relativity provided by Boyer \cite{boyer_1965}, were originally developed for the Schwarzschild case \cite{1978A&A....63..221A} and later generalized for the Kerr metric \cite{1978A&A....63..209K}.  
Regardless, such formalism can be used for a general asymmetric stationary spacetime, for which the metric components can be written as
\begin{align}
    ds^2=g_{tt}dt^2+2g_{t\varphi}dtd\varphi+g_{rr}dr^2+g_{\theta\theta}d\theta^2+g_{\varphi\varphi}d\varphi^2.
    \label{eq:generalmetric}
\end{align}

The tori are constructed by considering circular fluid motion around the rotation axis of the central compact object, thus the fluid four-velocity of any fluid particle can be written as
\begin{equation}
u^\mu=u^t(\eta^\mu+\Omega \xi^\mu),
\label{eq:circularU}
\end{equation}
where $\eta^\mu=\delta^\mu_t$ and $\xi^\mu=\delta^\mu_\varphi$ are the Killing vectors of the background space-time, while $\Omega\dot{=}\frac{u^\varphi}{u^t}$. 
By defining the specific angular momentum as the ratio $l=-\frac{u_\varphi}{u_t}$, we find the relations
\begin{equation}
    l=-\frac{g_{t\varphi}+g_{\varphi\varphi}\Omega}{g_{tt}+g_{t\varphi}\Omega},
    \qquad
    \Omega=-\frac{g_{t\varphi}+g_{tt}l}{g_{\varphi\varphi}+g_{t\varphi}l}.
    \label{eq:omegal}
\end{equation}
For normalized $u^\mu$, the four-acceleration of the fluid can be written as
\begin{equation}
a_\mu=\partial_\mu|\ln(u_t)|-\frac{\Omega}{1-\Omega l}\partial_\mu l .
\label{ac}
\end{equation}

Considering now a perfect fluid with rest-mass density $\rho$, specific enthalpy $h$ and pressure $p$, for which the stress-energy tensor reads
\begin{align}
    T_{\mu\nu}=\rho h u_\mu u_\nu +pg_{\mu\nu},
\end{align}
it is possible to write the fluid Euler equations:
\begin{equation}
-\frac{1}{\rho h}\partial_\mu p=\partial_\mu|\ln(u_t)|-\frac{\Omega}{1-\Omega l}\partial_\mu l   .
\label{euler}
\end{equation}
Assuming the fluid to have a barotropic equation of state, these equations can be integrated by summoning the von Zeipel's theorem \cite{1971AcA....21...81A,boyer_1965,Komissarov:2006nz}. 
The integral is performed from the outermost surface of the torus, where $p=0$ and $l=l_{\rm in}$, arriving at
\begin{equation}
    \mathcal{W}(r,\theta)-\mathcal{W}_{in}=\ln{|u_t|}-\ln{|(u_t)_{in}|}-\int_{l_{in}}^l\frac{\Omega dl'}{1-\Omega l'} \,,
    \label{eq:withl}
\end{equation}
where the potential $\mathcal{W}$ is defined as
\begin{equation} 
    \mathcal{W}(r,\theta)-\mathcal{W}_{in}\dot{=}-\int_0^p \frac{dp'}{\rho h} \, ,
    \label{defW}
\end{equation}
and $\mathcal{W}_{\rm in}$ is the potential calculated at the outermost surface of the torus. 
The potential $\mathcal{W}(r,\theta)$ holds information about the torus' rest-mass density and pressure distributions, specially in the case where the fluid's specific angular momentum distribution is set to be constant, for the integral at the right-hand side of Eq.~(\ref{eq:withl}) vanishes. 
In this special case, one can write $\mathcal{W}(r,\theta)$ explicitly 
\begin{align}
    \mathcal{W}(r,\theta)=\ln\left(\frac{g_{t\varphi}^2-g_{tt}g_{\varphi\varphi}}{g_{\varphi\varphi}+2l_0g_{t\varphi}+l_0^2g_{tt}} \right)^{\frac{1}{2}} \, ,
    \label{potentialW}
\end{align}
where $l_0$ is the constant specific angular momentum. 
An example of how the potential dictates the torus topology is given by considering the fluid to have a polytropic equation of state, $p=\kappa\rho^{\Gamma},$ where $\kappa$ is the polytropic constant and $\Gamma$ the polytropic index. In this case the rest-mass density distribution can be derived from $\mathcal{W}$ as follows
\begin{equation}
    \rho(r,\theta)=\left[ \left( \frac{\Gamma-1}{\kappa\Gamma}\right)[\exp(\mathcal{W}_{in}-\mathcal{W}(r,\theta))-1] \right]^{\frac{1}{\Gamma-1}}.
    \label{eq:density}
\end{equation}
For this distribution the isosurfaces of rest-mass density and pressure coincide with the isosurfaces of $\mathcal{W}(r,\theta)$. 
Rest-mass density values, given an equation of state and $\mathcal{W}$, can be rescaled ad hoc, as long as the assumption of a non-self-gravitating fluid holds. 
For instance, by choosing appropriate values of $\kappa$ and $\Gamma$ in Eq.~(\ref{eq:density}) it is possible to set a value of maximal rest-mass density in a given solution. 
Therefore, $\mathcal{W}$ holds most of the information of interest regarding the torus structure while being scale-independent. 
For this reason, we restrain ourselves to analyze mostly the potential hereafter.

Extrema of $\mathcal{W}(r,\theta=\pi/2)$ correspond to the locations of cusps and centers of the torus. 
At these locations the fluid moves also in Keplerian motion, for $\partial_r\mathcal{W}(r,\theta=\pi/2)=0 \implies a_\mu=0$. 
Thus, the circular orbits' specific angular momentum profiles for the spacetime to be analyzed determine how many cusps and centers would a torus solution be endowed with. 

These circular orbits in the equatorial plane of a given axisymmetric stationary spacetime can be addressed taking the same four-velocity as in Eq.~(\ref{eq:circularU}).
Defining the conserved quantities $E=-u_t$ and $L=u_\varphi$, representing the energy and angular momentum, the non-vanishing components of the four-velocity read
\begin{equation}
    u^t=-\frac{g_{t\varphi}L+g_{\varphi\varphi}E}{g_{t\varphi}^2-g_{tt}g_{\varphi\varphi}},\qquad u^t=\frac{g_{tt}L+g_{t\varphi}E}{g_{t\varphi}^2-g_{tt}g_{\varphi\varphi}}.
\end{equation}
Considering a space-like trajectory, $u^\mu u_\mu=-1$, the following effective potential can be defined:
\begin{equation}
    V_{\rm eff}\doteq g_{rr}\dot{r}^2=\frac{g_{\varphi\varphi}}{g_{t\varphi}^2-g_{tt}g_{\varphi\varphi}}\left(E-V_{+}\right)\left(E-V_{-}\right),
\end{equation}
where 
\begin{equation}
    V_{\pm}\doteq L\frac{g_{t\varphi}}{g_{\varphi\varphi}}\pm\frac{\sqrt{L^2(g_{t\varphi}^2-g_{tt}g_{\varphi\varphi})}}{g_{\varphi\varphi}}.
\end{equation}
Keplerian orbits are obtained when $V_{\rm eff}=\partial_r V_{\rm eff}=0$. 
These constraints allow us to find the angular velocity of such orbits, 
\begin{equation}
\label{eq:omega}
    \Omega_{\rm K}^{\pm}=\frac{-\partial_r g_{t\varphi}\pm\sqrt{(\partial_r g_{t\varphi})^2-\partial_r g_{tt}\partial_r g_{\varphi\varphi}}}{g_{\varphi\varphi}},
\end{equation}
which also allows us to access the specific angular momentum $l_{\rm K}^\pm$ of these geodesics through Eq.~(\ref{eq:omegal}). 
\footnote{As stated in Section 2, some orbits belonging to eigenvalues which usually correspond to the retrograde case (indexed as $l_{\rm K}^-$) have positive angular velocity, being effectively prograde orbits.}
The lower index $\rm K$ and upper index $+$ or $-$ indicates Keplerian orbits that are either prograde or retrograde. 

Given the profile $l_{\rm K}^\pm(r)$ and a constant specific angular momentum value $l_0$, the set of equatorial radii belonging to cusps or centers of a torus solution generated by $l_0$ will be given by $C=\{ r^* \,:\, l_{\rm K}^\pm(r^*)=l_0\}$. 
To distinguish cusps from centers, the stability of their corresponding Keplerian geodesics can be used. 
Stable orbits belong to the locations given by the set $S=\{r^*\, :\, \partial^2_r V_{\rm eff}(r^*)<0\}$. 
Thus, the set of centers is given by $C\cap S$, while the locations of cusps are given by $C\cap \bar{S}$.
The feasibility of multiple cusps or centers can be anticipated by analyzing the presence of local minima, maxima or discontinuities in $|l_K^{\pm}(r)|$. 
In the absence of discontinuities a local minimum makes it possible to have single-centered tori with one cusp. If one local minimum and one local maximum are present, double-centered tori, for which those centers are connected by a cusp, can be formed. 
When two local minima and one local maximum are present, double-centered tori endowed with two cusps can be formed. 
This structure is also possible for Keplerian specific angular momentum profiles endowed with only one local maximum and one local minimum, when discontinuities are observed. 
Examples of all of these cases are provided in Section 4.

As shown in \cite{Teodoro:2020kok}, spacetimes endowed with static orbits \cite{Collodel:2017end} can shelter static surfaces in the context of thick tori. While static orbits are circular geodesics in the equatorial plane, for which a particle remains static with respect to the ZAMO, static surfaces are their generalization for fluids. 
At these surfaces, the fluid is also at rest, $\Omega=0$, while inside it is moving in a prograde manner $\Omega>0$ and outside in a retrograde manner $\Omega<0$. 

The non-monotonicity of the metric functions is responsible for yet other peculiarities of the orbital velocity that arise in some of the solutions. 
Heeding the fact that $g_{tt}$ grows in the region bounded by the two static rings, if its slope is large enough, then the argument of the square root in Eq.~(\ref{eq:omega}) becomes negative and no inertial circular orbits can be realized. 
Note that where the argument becomes zero $\Omega_K^+=\Omega_K^-=\omega \rightarrow l=0$, i.e. the particle in circular orbit is a ZAMO. 
We should stress that even though there is a discontinuity in the existence of inertial circular orbits, a particle could still be in circular motion by applying the appropriate force in the radial direction. 

In order to track the existence of these surfaces given a torus solution with $l_0$, it is useful to define the {\it rest specific angular momentum} $l_r$ for which $\Omega=0$.
Applying this constraint to Eq.~(\ref{eq:omegal}), $l_r$ reads
\begin{equation}
    l_r=-\frac{g_{t\varphi}}{g_{tt}}.
\end{equation}
Thus, any torus distribution where $0>l_0>\min(l_r)$ will also house such surfaces. The intersection of these surfaces with the equatorial plane will be at the radii where $l_r(r)=l_0$.
Static orbits, being both Keplerian and having $\Omega=0$ are located at the radial positions $r_s$ for which the equality $l_K^-(r_s)=l_r(r_s)$ holds.

\section{Examples}

\begin{figure}[t]
\centering
\begin{subfigure}{.5\textwidth}
  \centering
  \includegraphics[width=1\linewidth]{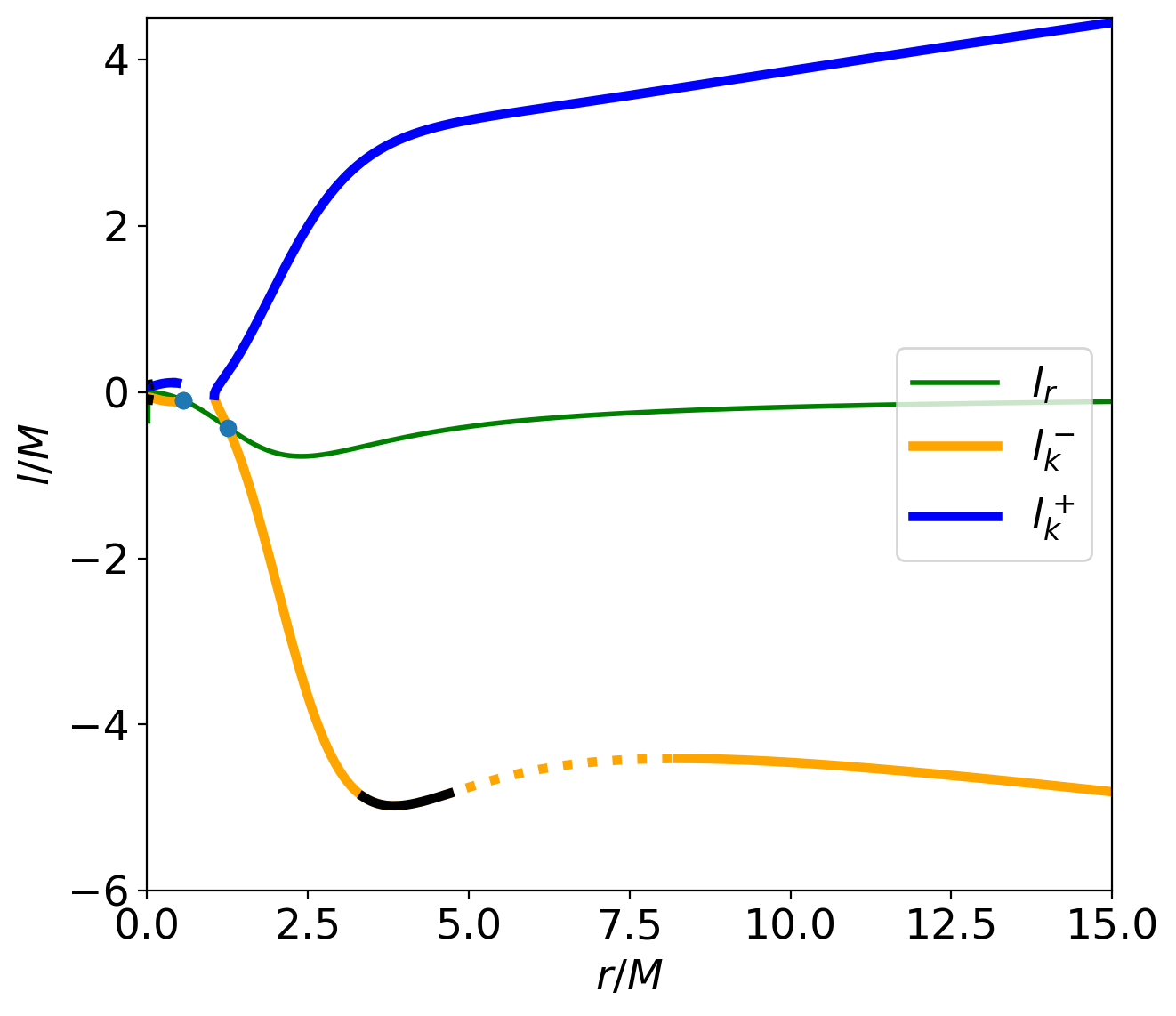}
  \caption{ }
       \label{fig:66}
\end{subfigure}%
\begin{subfigure}{.5\textwidth}
  \centering
  \includegraphics[width=1\linewidth]{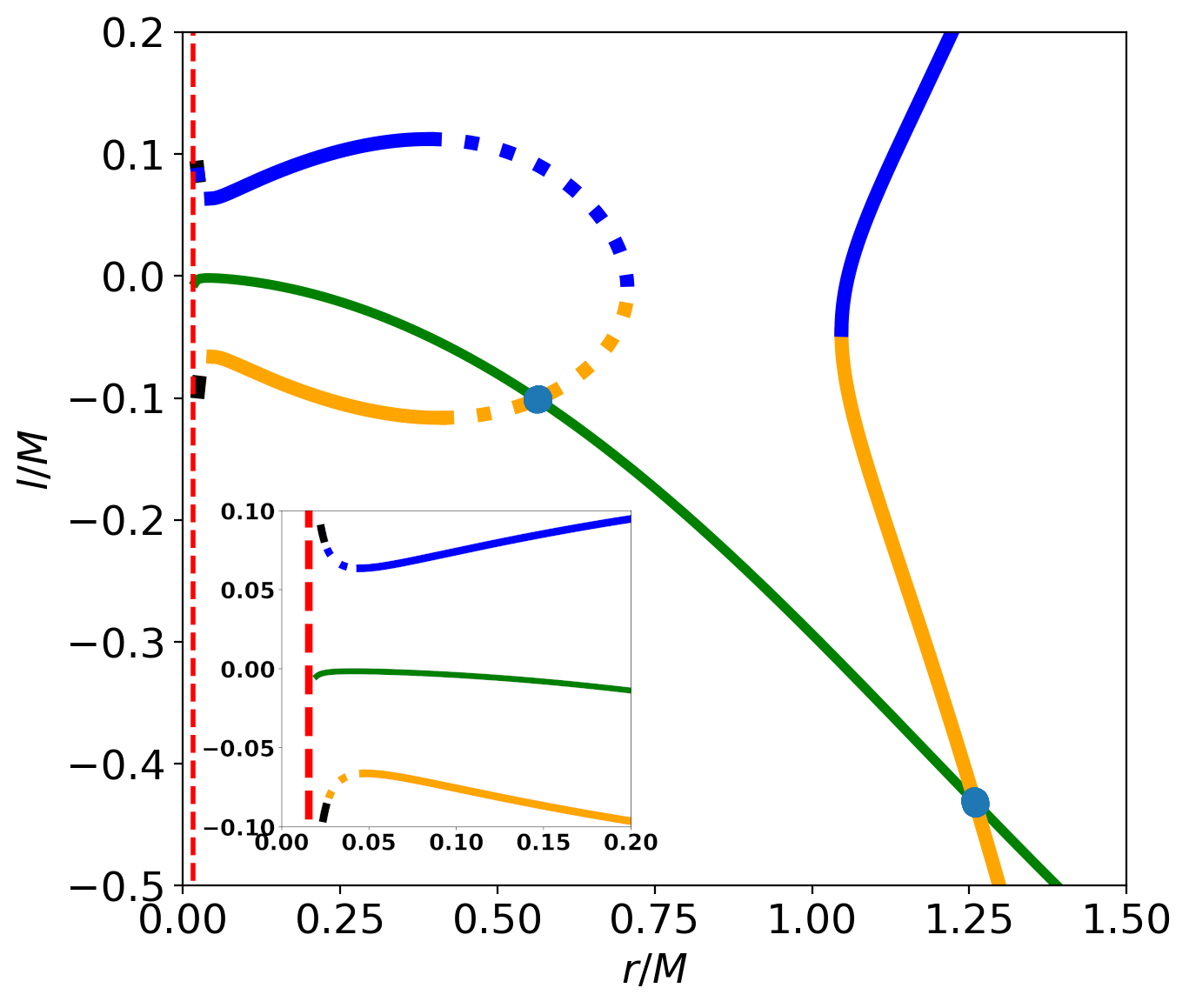}
 \caption{ }
       \label{fig:15}
\end{subfigure}
     \caption{Keplerian and rest specific angular momentum distribution on the equatorial plane of S1. Dotted lines represent unstable orbits, while black solid lines unbound ones. A zoom-in near the event horizon is provided in the right panel, where the dotted red line represents the normalized event horizon position.}
     \label{Fig:S2_l}
\end{figure}

\begin{figure}[h!]
\centering
\begin{subfigure}{.5\textwidth}
  \centering
  \includegraphics[width=1\linewidth]{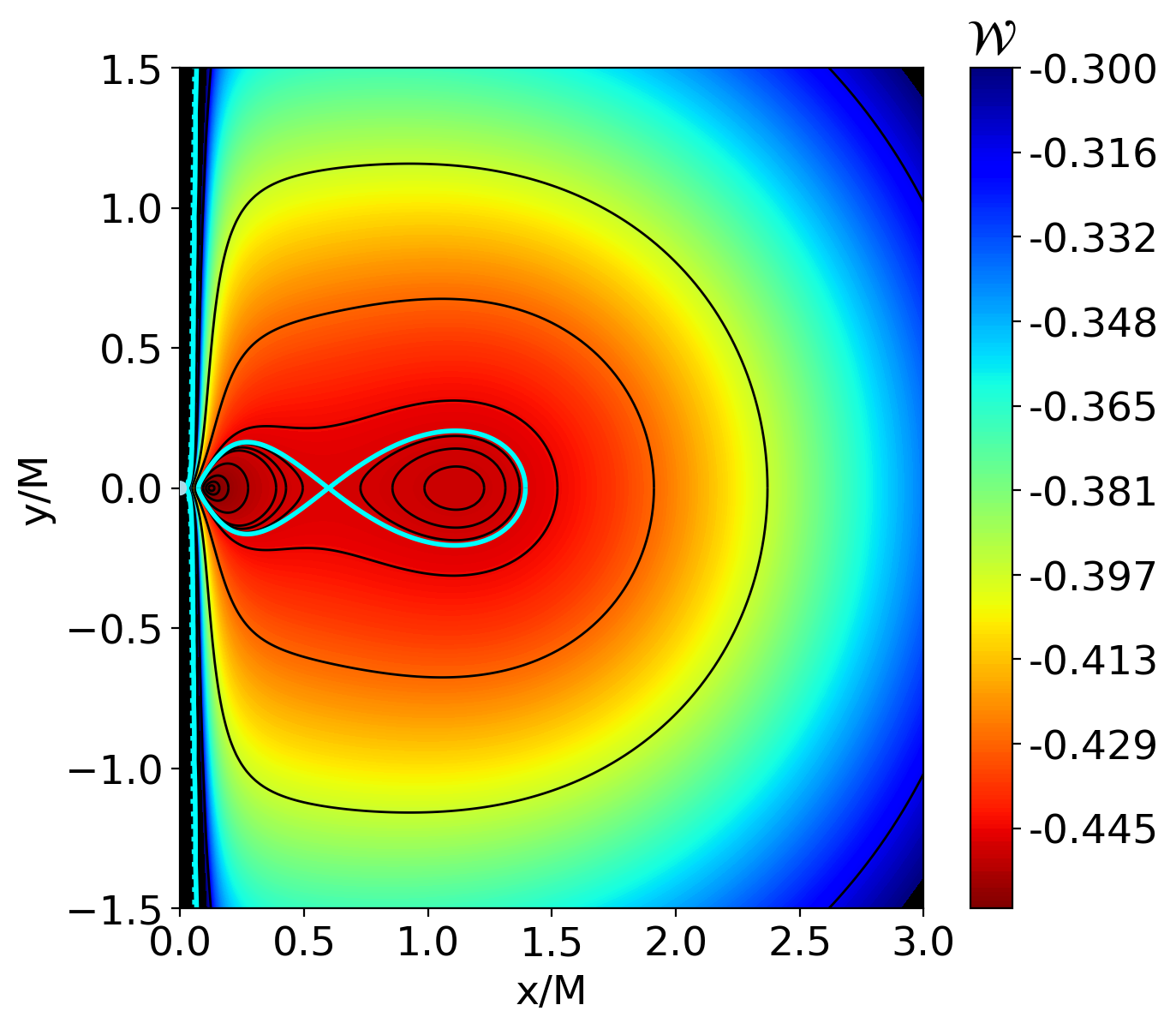}
  \caption{ }
       \label{fig:srbs}
\end{subfigure}%
\begin{subfigure}{.5\textwidth}
  \centering
  \includegraphics[width=1\linewidth]{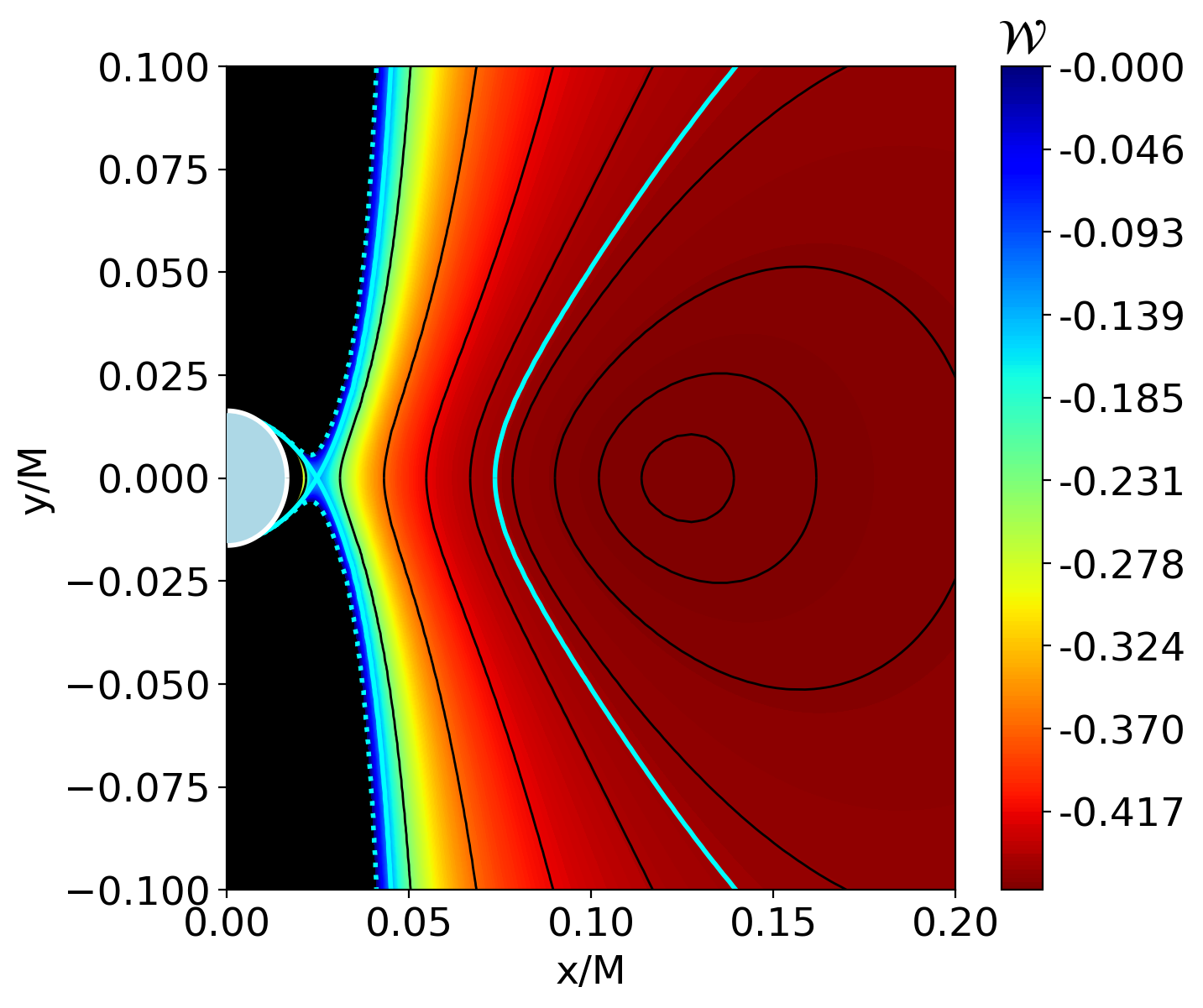}
 \caption{ }
       \label{fig:photonorbits}
\end{subfigure}
\begin{subfigure}{.5\textwidth}
  \centering
  \includegraphics[width=1\linewidth]{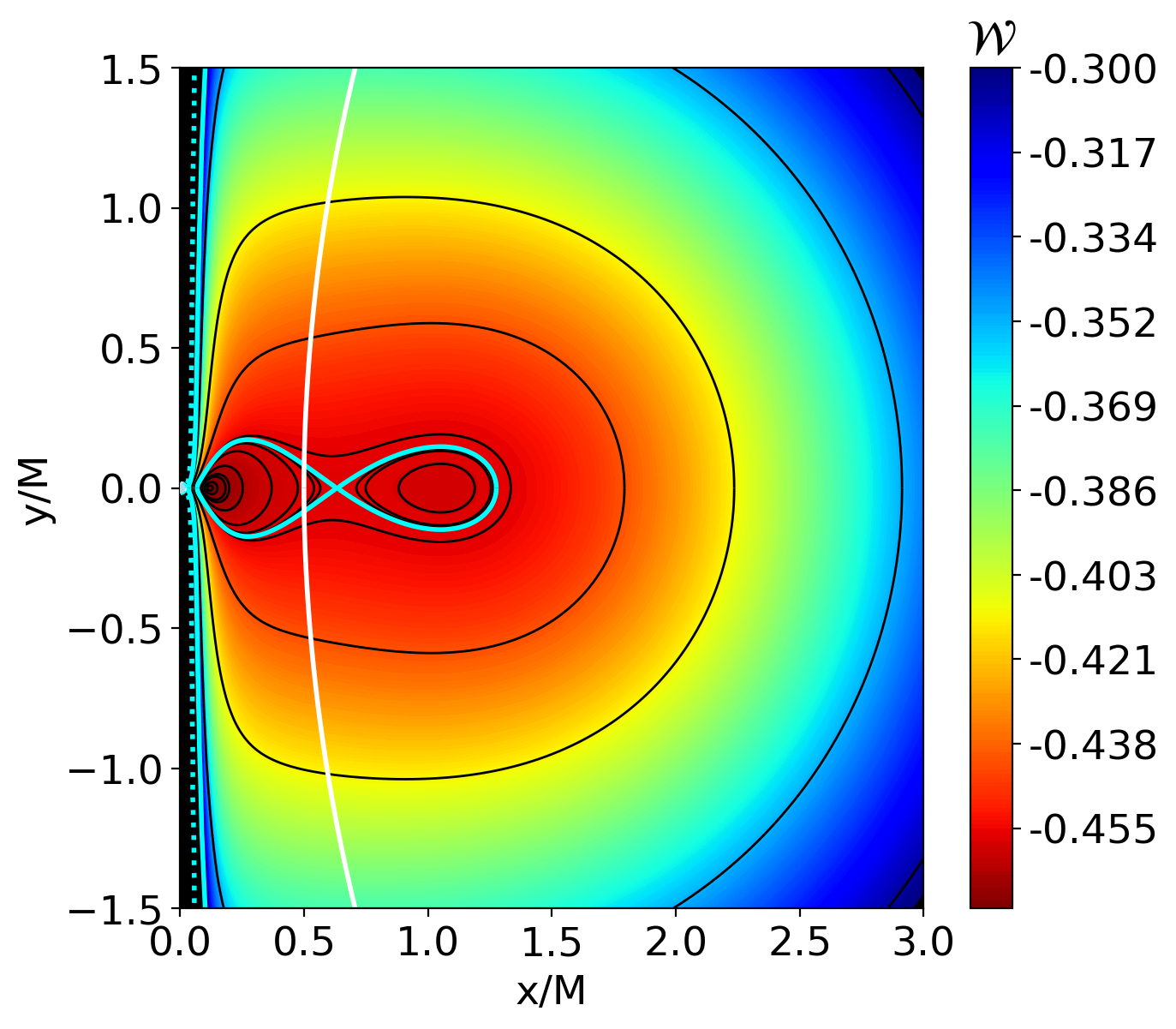}
  \caption{ }
       \label{fig:008minusfar}
\end{subfigure}%
\begin{subfigure}{.5\textwidth}
  \centering
  \includegraphics[width=1\linewidth]{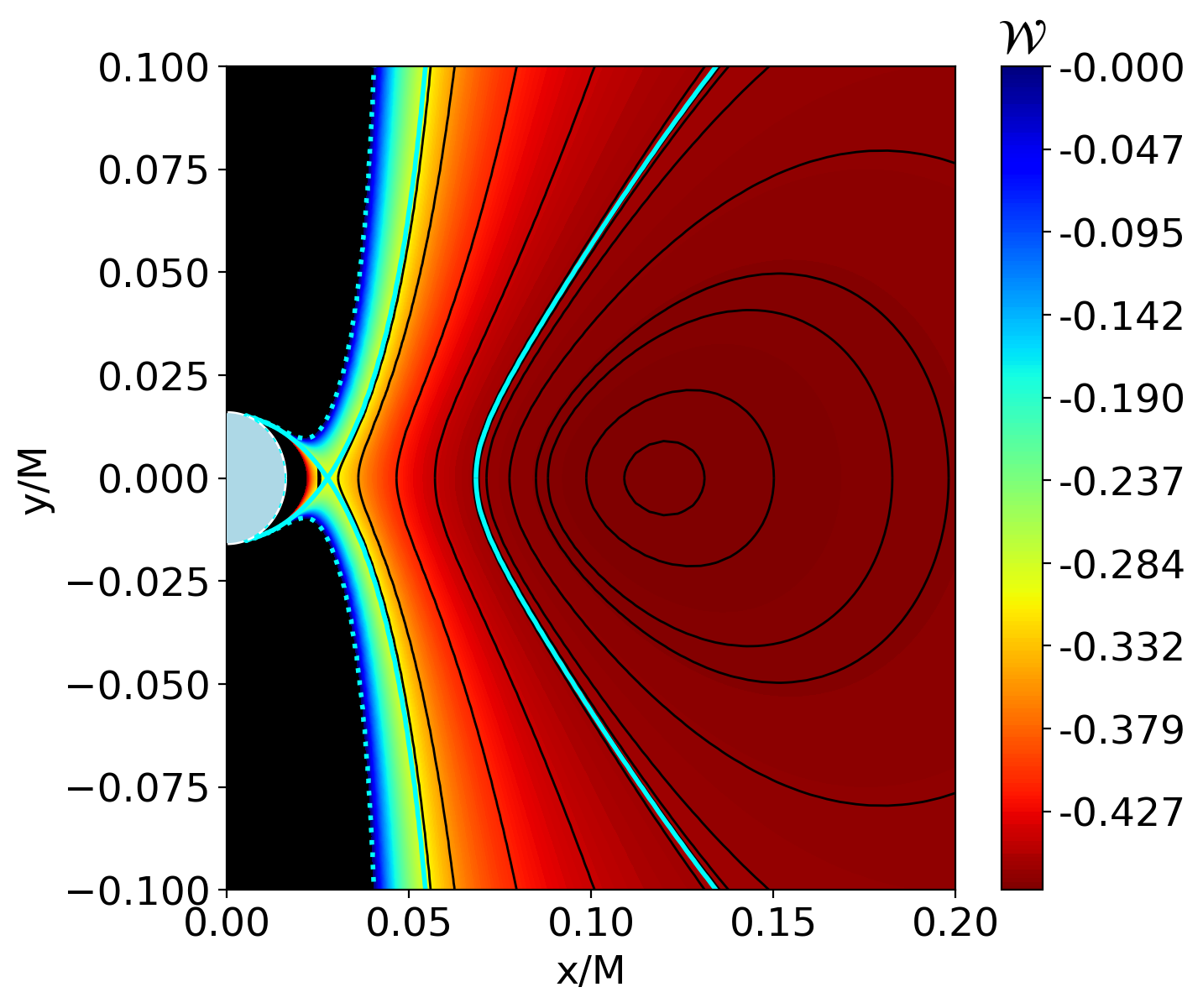}
  \caption{ }
       \label{fig:008minusclose}
\end{subfigure}%
     \caption{Potential for a torus around S1 endowed with $l_0=0.08M$ (panels a and b) and for a torus endowed with $l_0=-0.08M$ (panels c and d). 
     The bright solid lines represent the isosurfaces related to the cusps and the dotted blue lines correspond to the outermost isosurface of the solution, closed at infinity.}
     \label{Fig:S2_l008}
\end{figure}

In this section we shall consider the KBHsSH solutions depicted in Table \ref{Tab:1}, which represent cases of spacetimes endowed or not with a discontinuity in the Keplerian profile for angular velocities (DI) and/or static orbits (SO). 
These solutions were chosen since their Keplerian specific angular momentum profile presents prominent features that indicate interesting differences between tori housed by BS and by these solutions.
In fact, the Keplerian specific angular momentum around S1 has a DI while being endowed with two local minima for S2. 
In addition both solutions have static orbits, indicating the occurrence of static surfaces. 
The effects of these features on the construction of the Polish doughnuts  will be exemplified in this section.
The locations of these chosen solutions in the domain of existence of the KBHsSH are marked in Fig.~\ref{Fig:solutions}.

\begin{table}[h!]
\centering
\begin{tabular}{ |p{1cm}||p{1cm}|p{1cm}|p{1cm}|p{2cm}|p{1cm}| }
 \hline
 \multicolumn{6}{|c|}{KBHsSH solutions} \\
 \hline
 Label& $M$ & $r_h$ & $J$ & properties & $\omega$\\
 \hline
 S1 & $1.29$ & $0.02$ & $1.34$ & DI + SO & $0.800$\\
 S2 & $1.12$ & $0.07$ & $1.14$ & SO & $0.723$\\
 \hline
\end{tabular}
\caption{Labels, properties and relevant quantities of the two KBHsSH solutions taken as examples. The event horizon positions are not normalized by the mass of the solution.}
\label{Tab:1}
\end{table}

\begin{table}[h!]
\centering
\begin{tabular}{ |p{2cm}|p{1cm}|p{3cm}|p{3cm}|p{1cm}| }
 \hline
 \multicolumn{5}{|c|}{Torus solutions} \\
 \hline
 KBHsSH& $l_0/M$ & $r_{\rm center}/M$ & $r_{\rm cusp}/M$ & SS \\
 \hline
 S1 & $0.08$ & $[0.125 ,  1.114]$ & $ [0.025 , 0.598]$ & No \\
 S1 & $-0.08$ & $[0.119, 1.052]$ & $[0.028, 0.633]$ & Yes\\
 S1 & $-0.1$ & $[0.227, 1.059]$ & $[\emptyset, 0.567]$ & Yes \\
 S1 & $-0.431$ & $[1.260,\emptyset]$ & $[\emptyset,\emptyset]$ & Yes \\
 S2 & $0.455$ & $[0.311, 0.550]$ & $[0.114, 0.436]$ & No \\
 S2 & $-0.8$ & $[0.579, \emptyset]$ & $[0.124, \emptyset]$ & Yes \\
 \hline
\end{tabular}
\caption{Properties of the torus solutions reported. 
Positions of the cusps and centers can have two values, for multiple cusps and centers can be found. 
In the case of cusps, the first value corresponds to the innermost cusp connecting the tori with the event horizon, while the second corresponds to a cusp that connects two centers. 
"SS" refers to the absence or presence of a static surface in the solution. 
All values are normalized by the mass of the background solution.}
\label{Tab:2}
\end{table}

\begin{figure}[h!]
\centering
\begin{subfigure}{.5\textwidth}
  \centering
  \includegraphics[width=1\linewidth]{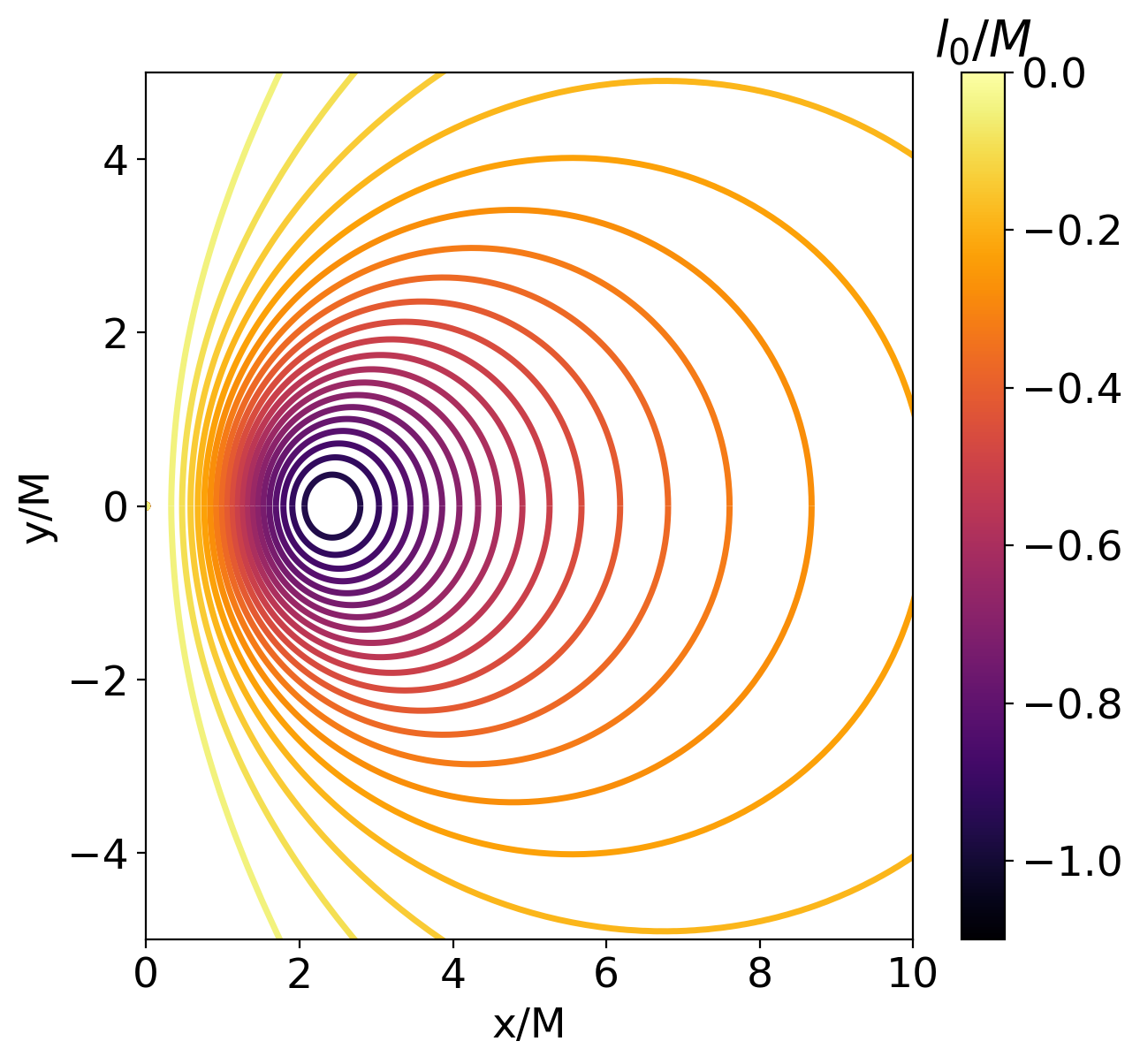}
  \caption{ }
       \label{Fig:S2_SS}
\end{subfigure}%
\begin{subfigure}{.5\textwidth}
  \centering
  \includegraphics[width=1\linewidth]{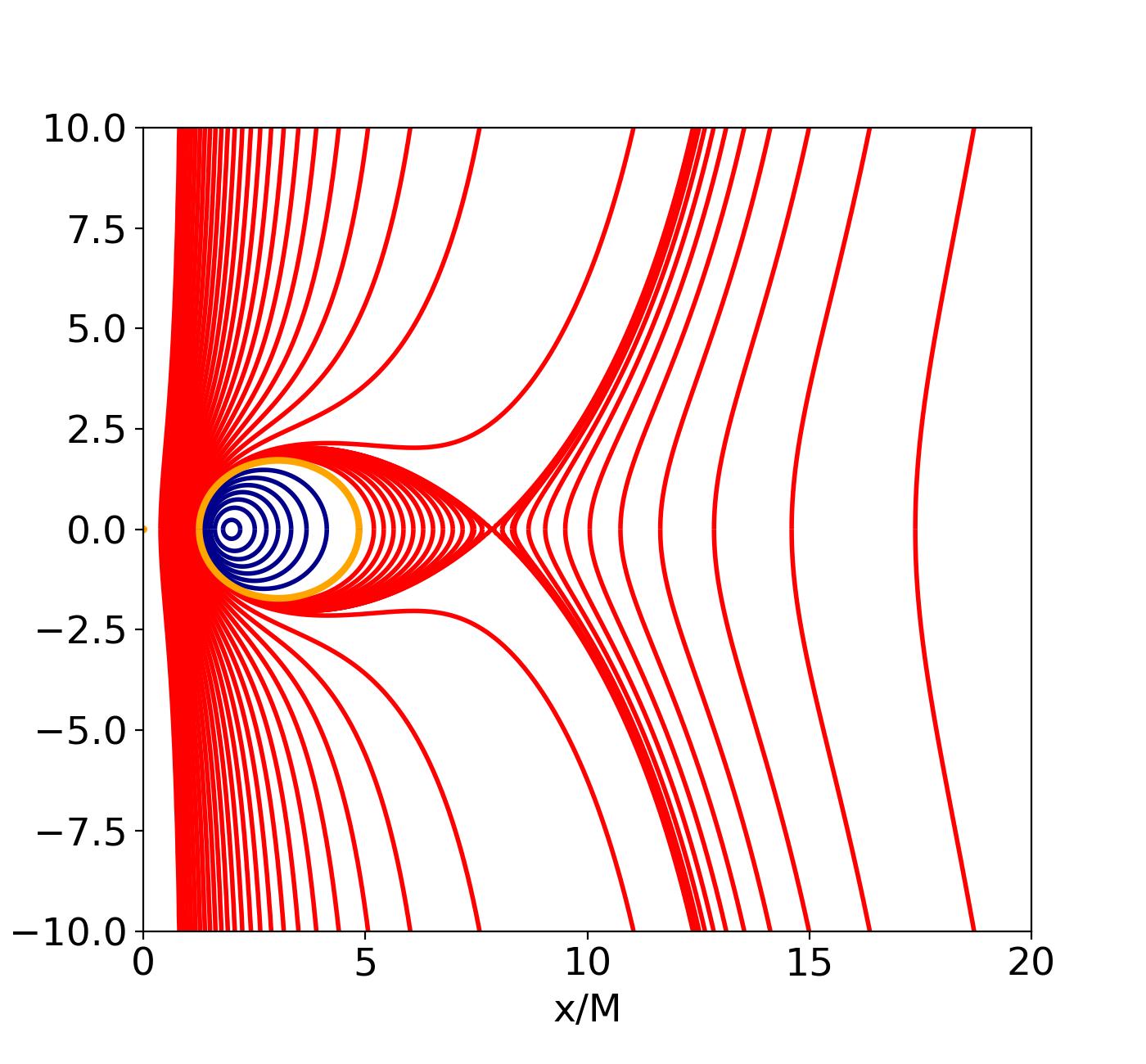}
 \caption{}
       \label{Fig:S2_SS_2}
\end{subfigure}
     \caption{(a) Static surfaces for tori sheltered by S1 as a function of $l_0$. 
     (b) Surfaces of constant angular velocity for a torus endowed with a static surface around S1 and constant specific angular momentum $l_0=-0.431M$. 
     Blue solid lines refer to positive values of $\Omega$, red lines to negative values of $\Omega$ and the yellow line to the static surface, i.e. $\Omega=0$.}
     \label{Fig:ss_vonzeipel}
\end{figure}

At first we shall consider the solution S1 (see Table \ref{Tab:1}), endowed with two static rings and a discontinuity in the specific angular momentum distribution for Keplerian orbits. 
In Fig.~\ref{Fig:S2_l} such a distribution is found together with the rest specific angular momentum profile. 
It can be seen that the equality $l_r=l_{\rm K}^{-}$ is satisfied in two radial positions,
namely, $r_s^{in}=0.567M$ and $r_s^{out}=1.260M$, where the indexing ``$\rm in$" (innermost) and ``$\rm out$" (outermost) is necessary in this situation, since two static rings exist. 
On the other hand, it can be seen that the innermost static ring is unstable, while the outermost is stable. 

Regarding the construction of tori, by avoiding the discontinuous region, Polish doughnuts similar to the ones found for rotating BSs emerge (\cite{Teodoro:2020kok} see section 5.2.2.). 
This means that prograde single-centered tori and retrograde single-centered or double-centered tori are accomplishable, as well as static surfaces. 
The novelties this solution brings can then be better illustrated by taking constant specific angular momenta $l_0$ that have small enough absolute value to reach the discontinuity of $l_{\rm K}^{\pm}$. 
We provide examples of tori with  $l_0=0.08M$, $l_0=-0.08$, $l_0=l_r(r_{s}^{\rm in})$ and $l_0=l_r(r_{s}^{\rm out})$.

In Fig.~\ref{Fig:S2_l008} a torus produced with the specific angular momentum $l_0=0.08$ around S1 is depicted.
This choice of $l_0$ produces a torus that is not only double-centered but also endowed with two cusps, a feature not yet found for Polish doughnuts, to our knowledge. 
One cusp connects the two centers, as typically happens for rotating BSs, but another cusp appears that connects the torus with the event horizon of the solution. 
It is also interesting to note that previously {\it prograde} doubled-centered tori were only found for BSs with winding number $k>3$ \cite{Meliani:2015zta}.
In contrast, for KBHsSH, $k=1$ is sufficient to produce such configurations.

In order to estimate the ratio
of the rest mass density at the inner center and at the outer center, we consider a polytropic fluid obeying $p=\kappa\rho^\Gamma$. 
In this case the ratio reads
\begin{align}
\frac{\rho^{(1)}}{\rho^{(2)}}={\Big(\frac{\exp({-\mathcal{W}^{(1)}})-1}{\exp({-\mathcal{W}^{(2)}})-1}\Big)}^{1/(\Gamma-1)} \, ,
\end{align}
with the superscripts $(1)$ and $(2)$ corresponding to the quantities evaluated at the inner and outer torus centers, respectively.
Taking the polytropic index to be $\Gamma=4/3$, we report that ${\rho^{(1)}}/{\rho^{(2)}}=1.09$, meaning that both centers ought to have similar densities, which contrasts with the results for BSs, where the innermost center was typically considerably denser than the outermost. 

Negative non-null specific angular momenta $l_0$ greater than $\min({l_r(r)})=-0.774M$ generate torus solutions endowed with a static surface, that reduces to a circle around the equatorial plane for $l_0=\min({l_r(r)})$. 
In Fig.~\ref{Fig:S2_SS}, a scheme with some examples of such surfaces for different torus specific angular momenta $l_0$ around S1 can be found.

Similarly to the case for $l_0=0.08M$, a double-centered torus endowed with two cusps is found for $l_0=-0.08M$. 
But now in contrast to the positive specific angular momentum solution a static surface appears. 
Thus the torus region containing the innermost center and the cusp moves in a retrograde manner, while the region containing the outermost cusp and center moves in a prograde manner.

\begin{figure}[h!]
\centering
\begin{subfigure}{.5\textwidth}
  \centering
  \includegraphics[width=1\linewidth]{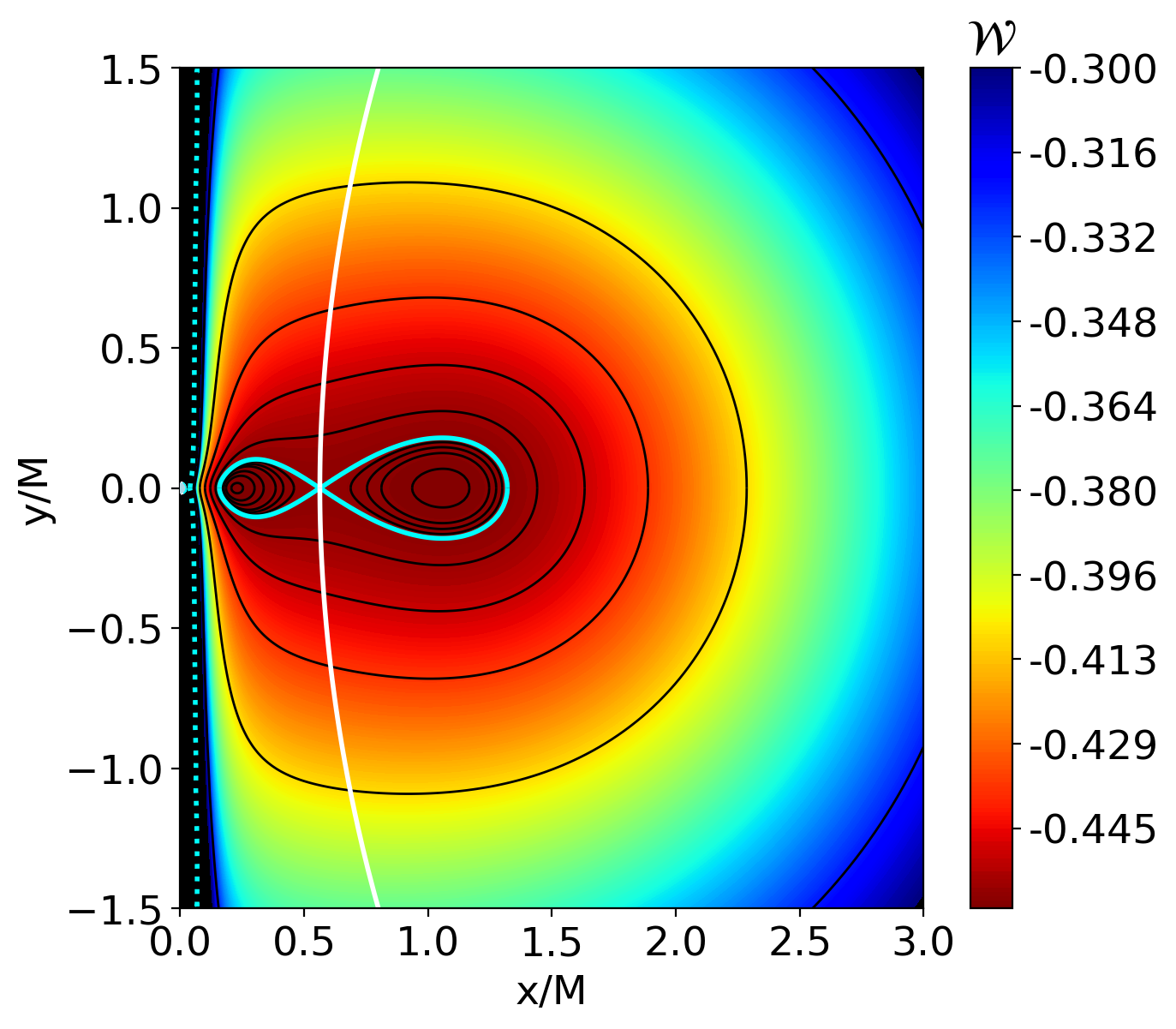}
  \caption{ }
       \label{fig:S2cusp}
\end{subfigure}%
\begin{subfigure}{.5\textwidth}
  \centering
  \includegraphics[width=1\linewidth]{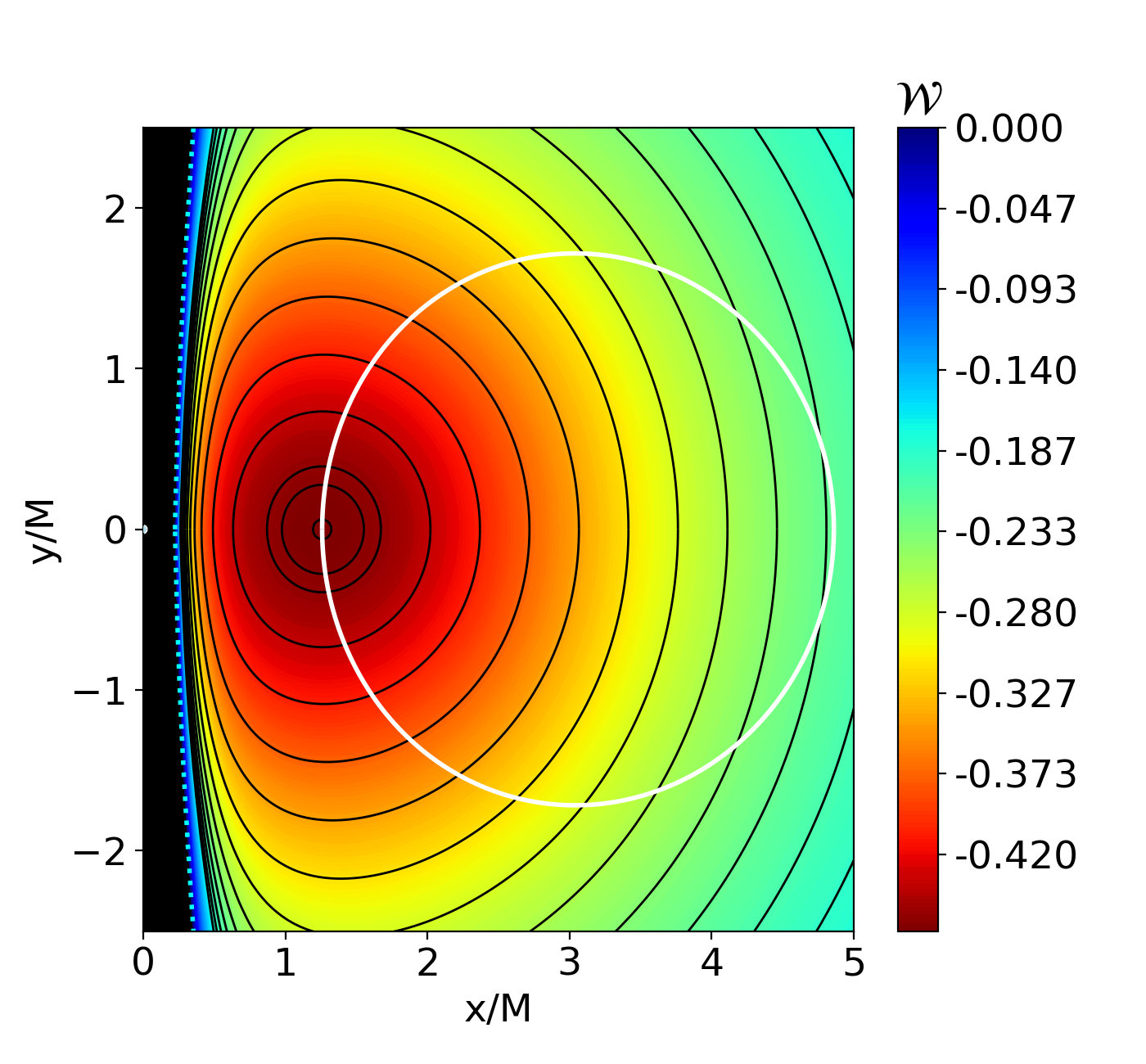}
 \caption{ }
       \label{fig:S2center}
\end{subfigure}
     \caption{Potential for tori around S1 endowed with $l_0=-0.100M$ (a) and $l_0=-0.431M$ (b). 
     The bright solid lines represent the isosurfaces related to the cusps, when existent, and the dotted blue lines the outermost isosurface of the solution, closed at infinity. 
     The static surface is plotted with a solid white line.}
     \label{Fig:S2_lminus}
\end{figure}

\begin{figure}[h!]
\centering
\begin{subfigure}{.5\textwidth}
  \centering
  \includegraphics[width=1\linewidth]{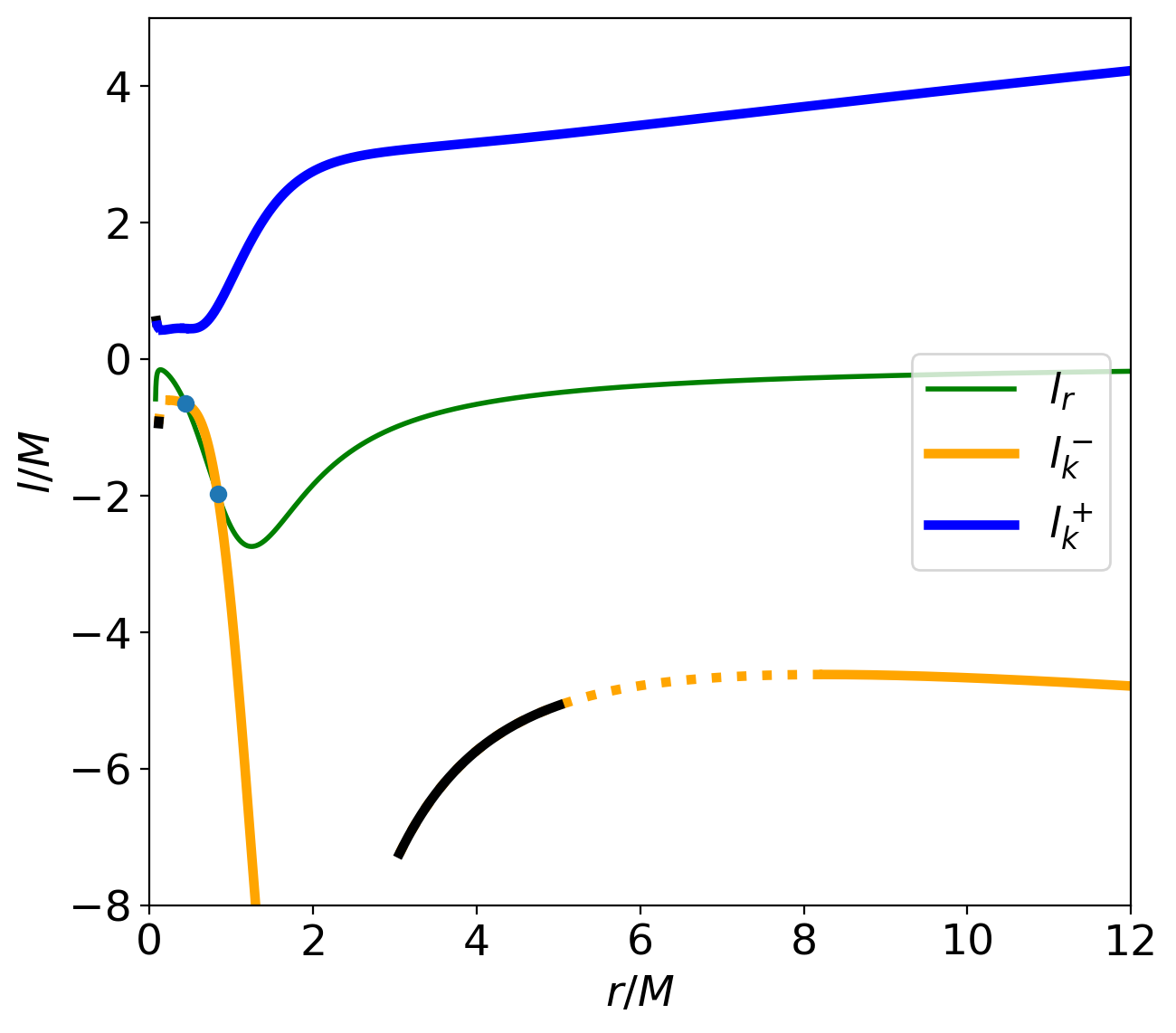}
  \caption{ }
       \label{fig:S3_l_a}
\end{subfigure}%
\begin{subfigure}{.5\textwidth}
  \centering
  \includegraphics[width=1\linewidth]{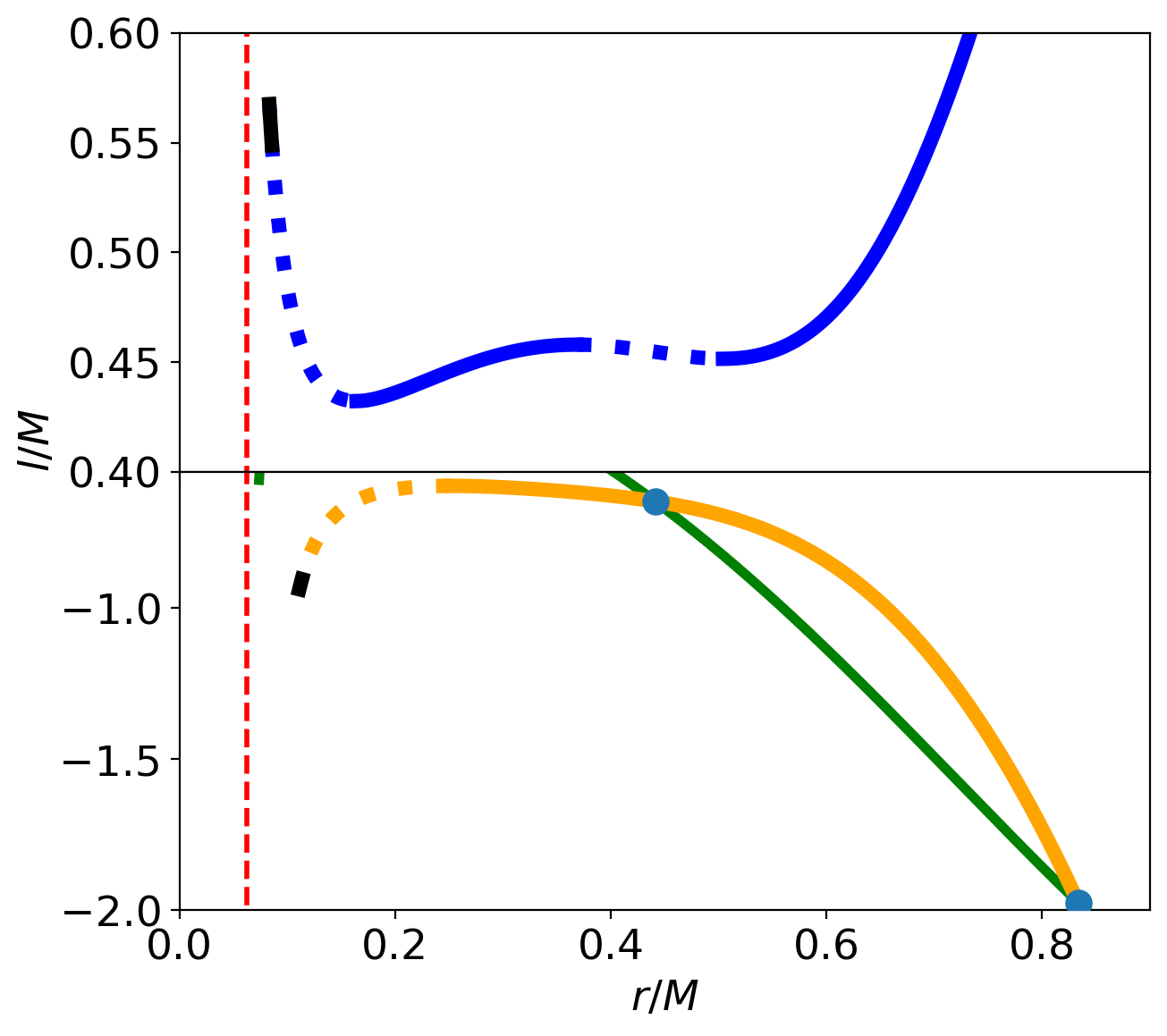}
 \caption{ }
       \label{fig:S3_l_b}
\end{subfigure}
     \caption{Keplerian and rest specific angular momentum distribution on the equatorial plane of S2.  
     Dotted lines represent unstable orbits, while black solid lines unbound ones. 
     A zoom-in near the event horizon is provided in the right panel, where the dotted red line represents the normalized event horizon position.}
     \label{Fig:S3_l}
\end{figure}

In the special cases when  $l_0=l_r(r_s)$, either a cusp or a center of the torus solution will be at a point of the static surface, depending on the stability of the static ring. 
As an example, we take $l_0=l_r(r_{s}^{\rm in})=-0.100M$, depicted in Fig.~\ref{fig:S2cusp}. 
The produced solution is endowed with two centers connected by a cusp. 
Such a cusp sits at the unstable Keplerian orbit that coincides with the static ring. 
The static surface of this solution separates the two sectors of the double-centered torus, where the innermost center is the only center in retrograde motion. 
The absence of an innermost cusp makes this solution similar to some solutions found for rotating BSs. 
But a contrasting feature should also be noticed. The static surface of these doubled-centered tori around BSs is typically small and nearby the innermost center and can not reach the cusp. 
Again the estimated ratio of the densities of the two centers for a polytropic equation of state is close to one, $\rho^{(1)}/\rho^{(2)}=1.01$.

In contrast, by choosing $l_0=l_r(r_{s}^{\rm out})=-0.431M$, depicted in Fig.~\ref{fig:S2center}, a single-centered cuspless torus is produced, for which the center is at the stable static orbit. 
To illustrate the angular velocity distribution of solutions like this, endowed with a static surface, we also provide a plot of the von Zeipel surfaces, i.e. the surfaces of constant angular velocities in Fig.~\ref{Fig:S2_SS_2}. 
Most of the pattern of the von Zeipel surfaces simply follows the well-known pattern for Kerr black holes,
with spheroidal surfaces close to the horizon, paraboloidal surfaces in the vicinity of the symmetry axis and cylindrical surfaces otherwise, except for the self-interacting surface at the transition radius \cite{Chakrabarti:1990}.
However, for solutions with a static surface, inside the cylindrical region another spheroidal region arises that contains the static surface and the co-rotating fluid.
Of course, the presence of this new cylindrical region entails the presence of a further transition radius with associated self-intersecting surface.
The figure highlights only these new features due to the static surface.
  
Moving to solution S2, the specific angular momentum is depicted in Fig.~\ref{Fig:S3_l}. 
No discontinuity in the Keplerian distribution is found, except for the branch of retrograde superluminal orbits,
which is also characteristic of some rotating BSs. 
Once more, for relatively large values of $|l_0|$, tori similar to the ones found for rotating BSs emerge, while by choosing lower values of $|l_0|$ cusps connecting the tori to the event horizon are produced. 

\begin{figure}[h!]
\centering
\begin{subfigure}{.5\textwidth}
  \centering
  \includegraphics[width=1\linewidth]{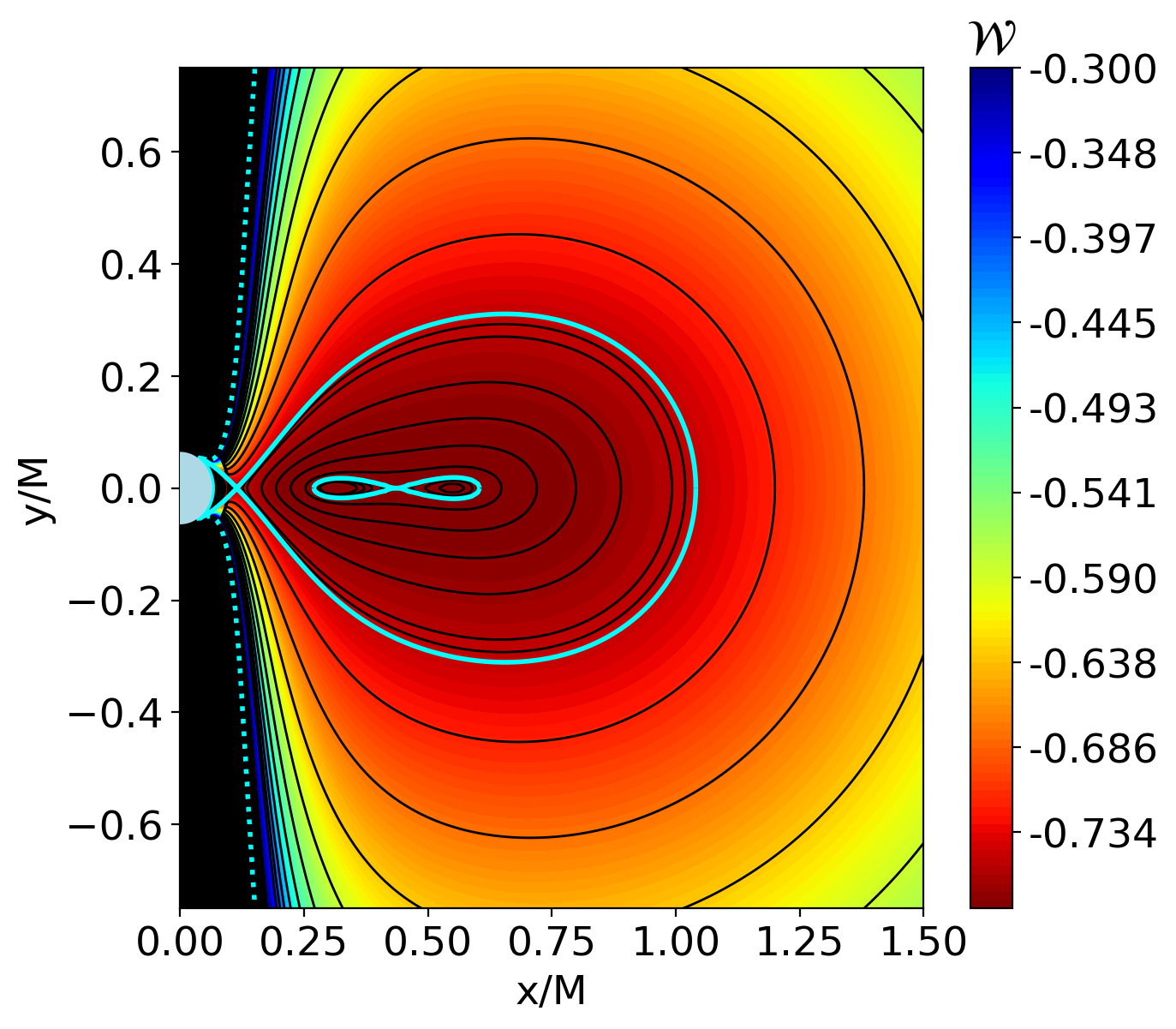}
  \caption{ }
       \label{fig:S3plus0455}
\end{subfigure}%
\begin{subfigure}{.5\textwidth}
  \centering
  \includegraphics[width=1\linewidth]{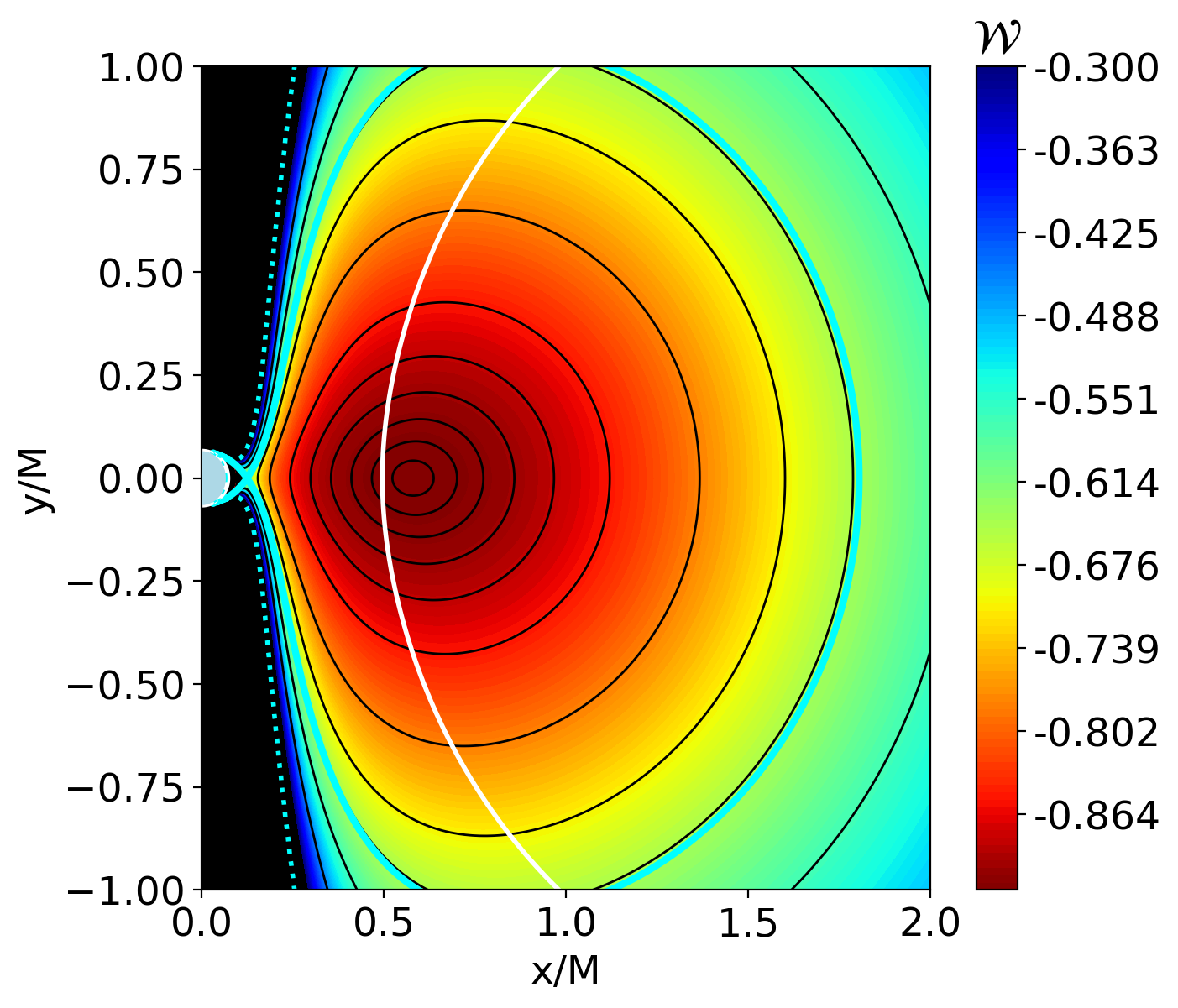}
 \caption{ }
       \label{fig:S3minus08}
\end{subfigure}
     \caption{Potential for tori around S2 endowed with $l_0=0.455M$ (a) and $l_0=-0.8M$ (b). 
     The bright solid lines represent the isosurfaces related to the cusps, and the dotted blue lines correspond to the outermost isosurface of the solution, closed at infinity. 
     The static surface is plotted with a solid white line, when existent.}
     \label{fig:S3tori}
\end{figure}

As seen in Fig.~\ref{fig:S3_l_b}, the specific angular momenta of the prograde Keplerian orbits show a local maximum and two local minima near the event horizon, meaning that regardless of the absence of a discontinuity, double-centered tori with two cusps can be formed. 
As an example, we choose $l_0=0.455M$ for the solution depicted in Fig.~\ref{fig:S3plus0455}. 
Here the two centers are closer than the ones in the previous solutions. 
Furthermore, 
$\rho^{(1)}/\rho^{(2)}=1.00008$.
In contrast, such topologies are not achievable in the retrograde case in the absence of a discontinuity.
Yet single-centered tori with a cusp and a static surface can be formed, which is the case, for instance, for $l_0=-0.8M$ (Fig.~\ref{fig:S3minus08}).

\section{Conclusions}

By approaching thick tori around KBHsSH solutions with a constant specific angular momentum Polish doughnut model, we reported in this publication possible non-mainstream geometries that these fluid configurations can be endowed with. 
Characteristics similar to the ones already observed for rotating BSs, but absent for Kerr BHs, were found also for KBHsSH, such as double-centered tori, with the two centers connected by a cusp.
We also found several important differences between the two types of objects.
The ratio between the densities of the centers were found to be close to one for KBHsSH, while for BSs the innermost center is typically considerably denser than the outermost.
Furthermore, an extra cusp, now connecting the fluid to the event horizon was also found for KBHsSH, a feature not observed before in any spacetime to our knowledge. 
With the examples provided, we found that these double-centered configurations with two cusps are accomplishable both in the retrograde and prograde case when a discontinuity on the Keplerian specific angular momentum distribution exists.
On the other hand, these torus geometries can also be achieved for some cases when such a discontinuity is absent, but only in the prograde case. 

Another appealing feature of the Polish doughnuts around KBHsSH is that they can also shelter static surfaces. 
These surfaces, a generalization of static orbits for non-geodesic fluid configurations, have vanishing angular velocity with respect to the ZAMO. 
When existent, the fluid moves in a prograde manner inside the surface and in a retrograde manner outside. 
Differently than static surfaces for BSs, in the context of KBHsSH these surfaces can be housed by single-centered tori with a cusp. 
In the context of double-centered tori, they are also accomplishable, but are no longer found in the vicinity of the innermost center and can even contain the cusp connecting the two centers. 

In order to investigate the role of this extra cusp as well as the evolution of tori with static surfaces, further simulations would be useful. 
The addition of magnetic fields in the analytical solution as well as the investigation of different specific angular momentum distributions are also fruitful next steps for the solutions here presented. 

\section{Acknowledgements}

We gratefully acknowledge support by the DFG funded
Research Training Group 1620 ``Models of Gravity''.
LGC and DD would like to acknowledge support via an
Emmy Noether Research Group funded by the DFG
under Grant No. DO 1771/1-1.
PN is partially supported by the Bulgarian NSF Grant  KP-06-H38/2.
SY acknowledges support by the University of Tuebingen and the Bulgarian
NSF Grant KP-06-H28/7. 
The authors would also like to acknowledge networking 
support by the COST Actions CA15117, CA16104, and CA16214.

\bibliographystyle{elsarticle-num}
\bibliography{biblio.bib}







\end{document}